\documentclass{amsart}

\usepackage{amssymb}
\usepackage{enumerate, xspace}
\usepackage{dsfont}
\usepackage{afterpage}
\usepackage{changebar}
\usepackage[dvips]{graphicx}
\usepackage{amsthm}
\usepackage{rotating}

\numberwithin{equation}{section}

\theoremstyle{definition}

\newcommand{\be}{\begin{equation}}
\newcommand{\ee}{\end{equation}}
\newcommand{\bea}{\begin{eqnarray}}
\newcommand{\eea}{\end{eqnarray}}
\newcommand{\bs}{\begin{split}}
\newcommand{\es}{\end{split}}

\begin{document}


\title[Reversible carrier-type transition]
  {Reversible carrier-type transition in gas-sensing oxides and nanostructures}
\author[Arulsamy et al.]{Andrew Das Arulsamy} \email{andrew.das.arulsamy@ijs.si}

\address{Jo$\check{z}$ef Stefan Institute, Jamova cesta 39, SI-1000 Ljubljana, Slovenia}

\author[]{Kristina Eler$\check{\rm s}$i$\check{\rm c}$}

\author[]{Martina Modic}

\author[]{Uro$\check{\rm s}$ Cvelbar}

\author[]{Miran Mozeti$\check{\rm c}$}

\keywords{Quantum adiabatic theorem and approximation; Degenerate energy levels; Time-dependent orthogonalization; Internal and external time scales}

\begin{abstract}
Despite many important applications of $\alpha$-Fe$_2$O$_3$ and Fe doped SnO$_2$ in semiconductors, catalysis, sensors, clinical diagnosis and treatments, one fundamental issue that is crucial to these applications remains theoretically equivocal$-$ the reversible carrier-type transition between $n$ and $p$-type conductivities during gas-sensing operations. Here, we give unambiguous and rigorous theoretical analysis in order to explain why and how the oxygen vacancies affect the $n$-type semiconductors, $\alpha$-Fe$_2$O$_3$ and Fe doped SnO$_2$ in which they are both electronically and chemically transformed into a $p$-type semiconductor. Furthermore, this reversible transition also occurs on the oxide surfaces during gas-sensing operation due to physisorbed gas molecules (without any chemical reaction). We make use of the ionization energy theory and its renormalized ionic displacement polarizability functional to reclassify, generalize and to explain the concept of carrier-type transition in solids, and during gas-sensing operation. The origin of such a transition is associated to the change in ionic polarizability and the valence states of cations in the presence of (a) oxygen vacancies and (b) physisorped gas molecules. 
\end{abstract}

\keywords{Carrier-type transition, Gas-sensing oxides, Ionization energy theory, Ionic polarizability, Oxygen vacancies and interstitial defects}

\maketitle

\section{Introduction}

Carrier-type transitions ($n$- to $p$-type) in oxides can be achieved by doping Fe$^{2+,3+}$ into the $n$-type SnO$_2$. Interestingly, this Fe doped SnO$_2$ also behaves like an $n$-type semiconductor if the surface is exposed to oxygen molecules at higher temperatures.~\cite{gala} Note here that these transitions ($n \rightarrow p \rightarrow n$) have different causes (but same origin), in which, the former $n$- to $p$-type transition is entirely due to Fe doping. The latter $p$- to $n$-type transition is due to the physisorbed oxygen molecules onto the $p$-type surface (Fe-doped SnO$_2$ surface) at elevated temperatures. We will come to understand in the subsequent sections why these two transitions are non-trivially chemical and electronic in origin. For example, apart from temperature, the carrier-type transition can only be tuned by \\

$\textsc{X1}$: changing the chemical element compositions or their valence states (by creating defects) on the surfaces of sensors or

$\textsc{Y1}$: changing the chemical element compositions of gaseous (or its concentrations) that are in contact with the surfaces. \\

These manipulations will give rise to the corresponding changes to the electronic polarization of the surface atoms or ions, as well as the adsorbed neutral or charged gas molecules, which will eventually lead to the reversible carrier-type transition. Note here that higher temperatures give rise to more excited electrons (due to Fermi-Dirac statistics), which then enhances the conductivity due to electrons ($n$-type semiconductor).~\cite{gala} 

In fact, the first observation for the carrier-type transition due to $\textsc{Y1}$ was observed in Cr$_2$O$_3$ exposed to ethanol by Bielanski et al.~\cite{biel} On the other hand, $\alpha$-Fe$_2$O$_3$ is an $n$-type in pure form (without defects) that has been made to be a $p$-type semiconductor by deliberately removing the oxygen atoms from the bulk $\alpha$-Fe$_2$O$_3$ nanostructures [transition due to $\textsc{X1}$].~\cite{lee} Subsequently, these nanostructures show an $n$-type [$p$- to $n$-type transition due to $\textsc{X1}$ again] behavior when they are exposed to reductive ambient.~\cite{lee} Lee et al.~\cite{lee} proposed that oxygen ordering in the bulk is the cause for $p$- to $n$-type transition. Apparently, these results are different from the oxygen ion adsorption model that takes surface work function into account via the surface contact potential.~\cite{gur} For example, the bulk $\alpha$-Fe$_2$O$_3$ sample is an $n$-type semiconductor, which has been made to be a $p$-type after exposing it to high concentration of oxygen gas. This $n$- to $p$-type transition is associated to an increased adsorbed oxygen ions, an increased in band bending and a higher hole concentration.~\cite{gur} 

Subsequent exposure to CO (reductive ambient) have resulted in CO reacting with the adsorbed surface oxygen, thus releasing free electrons into the surface conduction band to give $n$-type conductivity.~\cite{gur} Clearly, the $n \rightarrow p \rightarrow n$ transitions observed by Gurlo et al.~\cite{gur} are entirely due to $\textsc{Y1}$, and are different from the $n \rightarrow p \rightarrow n$ transitions reported by Lee et al.~\cite{lee}, which are due to $\textsc{X1}$ only. In addition, the $n \rightarrow p \rightarrow n$ transitions discussed by Galatsis et al.~\cite{gala} are caused by both $\textsc{X1}$ and $\textsc{Y1}$. Therefore, it is essential to evaluate these electro-chemical carrier-type transitions due to $\textsc{X1}$ or $\textsc{Y1}$ or both explicitly, and microscopically (with quantum effects).  

However, there is only one semi-classical explanation available (based on the conduction, valence and impurity bands) on the carrier-types in semiconductors as given by Ashcroft and Mermin.~\cite{ash} These commonly accepted theoretical definitions for $n$-type and $p$-type semiconductors are based only on cations oxidation or valence states. For example, semiconductors with holes as charge carriers are known as $p$-types as a result of the dopants oxidation states, which are lower than the matrix intrinsic semiconductor. For instance, B doped Si (matrix) is a $p$-type and B$^{3+}$ here acts as an acceptor that attracts free holes from Si$^{4+}$. Likewise, As doped Si semiconductor is an $n$-type with electrons as charge carriers due to positively-charged As$^{5+}$ (acts as a donor)~\cite{ash} attracts free electrons from Si$^{4+}$. Apparently, these definitions cannot be applied directly for oxides that has oxygen concentration as an additional doping variable.      

Therefore, re-evaluation of our microscopic understanding of this carrier-type transition are not only crucial for conceptual understanding, but it is also essential for the advancement of numerous applications, such as in gas sensors, solar energy converters, water splitters, photocatalysts, chemical warfare agents (for defense), and for cancer-cell ablation, shape-memory effect and bio-imaging of cells.~\cite{gouma,pras,uros,gouma2,sunu,chau,bars,smit,kaur,chir,soo,cwang,mohr,nair,nair2,hua,xie,huber} In this work, however, we will not study exotic oxides such as cuprates and manganites, which have been discussed elsewhere with respect to oxygen-deficient normal state conductivity and ferromagnetism.~\cite{and1,and2} Apart from these applications, specially coated $\alpha$-Fe$_2$O$_3$ (for bio-stability) is also used in the fields of biomedical and bioengineering, such as for drug and protein delivery, and for cell separation and detoxification of biological fluids due to its magnetic properties.~\cite{po,gup,guo,rui,chin,jin} In this case, the existence of Fe$^{2+}$ and Fe$^{3+}$ in $\alpha$-Fe$_2$O$_{3-x}$ due to oxygen vacancies ($x$) may improve its magnetic properties as it resembles the ferrimagnetic Fe$_3$O$_4$, which also contains the mixed iron valence.~\cite{ver}

As pointed out earlier, the carrier-type transition is also closely related to oxygen vacancies on the solid surfaces [$\textsc{X1}$]. The existence of this association has been shown experimentally by Lee et al.~\cite{lee} Within the context of oxygen vacancies, Chen et al.~\cite{chen} have also reported the process of systematic removal of oxygen anions via diffusion from the $\alpha$-Fe$_2$O$_3$ nanowires and nanobelts. Creation of such vacancies in $\alpha$-Fe$_2$O$_3$ were possible due to an innovative and universal approach proposed by Cvelbar et al.~\cite{uros2} In this technique, the synthesis of $\alpha$-Fe$_2$O$_3$ nanowires and nanobelts were carried out by direct plasma oxidation of bulk Fe materials. This method was originally proposed by Mozeti$\check{\rm c}$ et al.~\cite{moz} to synthesize high-density niobium oxide nanowires. For more general and recent details on the properties of iron oxides, we refer the readers to a recent review by Chirita and Grozescu~\cite{chir} that covers the different polymorphs of the iron oxides as well as their electrochemical properties in the context of photoelectrochemistry. 

It is to be noted here that rigorous experimental investigations on the $p$- to $n$-type transition have been carried out by Vaidhyanathan et al.~\cite{vai}, and Murugavel and Asokan~\cite{muru} for chalcogenide glasses. They used the model developed by Kolobov~\cite{kolo} that is based on the modification of the charged defect states and dangling bonds to explain such carrier-type reversal as a result of Pb doping in Pb-Ge-Se-Te glasses. In addition to that, the effect of polarizability in this transition was also first proposed by Elliot and Steel.~\cite{ell} In this work, however, we will need to exclude this material (chalcogenides) because all the elements in this class of materials are multi-valent, for example, Pb$^{2+,4+}$, Ge$^{2+,4+,4-}$, Se$^{2+,4+,6+,2-}$ and Te$^{2+,4+,6+,2-}$. Therefore, it is unlikely for us to narrow down the origin of carrier-type transition in chalcogenides based on the IET. In other words, the uncertainties from the IET analysis will be too large for one to make any sense. 

The paper is organized as follows. Theoretical details on the IET are only briefly discussed where the details can be found elsewhere. A thorough and rigorous analysis are given on the carrier-type transition in both traditional, oxide semiconductors, and the $n$- and $p$-type carrier classifications. Subsequently, we apply our refined classification to explain the reversible carrier-type transition in $\alpha$-Fe$_2$O$_3$ and Fe doped SnO$_2$ due to $\textsc{X1}$. Prior to conclusions, further analysis are carried out to explain the carrier-type transition during sensor operation [due to $\textsc{Y1}$ in the presence of $\textsc{X1}$]. We conclude that all of our theoretical analysis are in agreement with the experimental results reported in Refs.~\cite{gala,gur,lee} without any violation and with high-level self consistency.     

\section{Theoretical details}

The theoretical concept described here is based on the ionization energy theory (IET) formalism~\cite{and1,and3,and4} where the polarizability of cations and anions can be obtained from the renormalized ionic displacement polarizability equation~\cite{and8}, 

\begin {eqnarray}
\alpha_{d} = \frac{e^2}{M}\bigg[\frac{\exp[\lambda(E_F^0 - \xi)]}{(\omega_{\rm{ph}}^2 - \omega^2)}\bigg], \label{eq:2}
\end {eqnarray}  
  
where $\omega_{\rm{ph}}$ is the phonon frequency of undeformable ions, $1/M = 1/M^+ + 1/M^-$ in which, $M^+$ and $M^-$ are the positively and negatively charged ions due to their different polarizabilities. Here, $\lambda = (12\pi\epsilon_0/e^2)a_B$, $a_B$ is the Bohr radius of atomic hydrogen, $e$ and $\epsilon_0$ are the electronic charge and the permittivity of space, respectively. Recent examples on the application of IET are reported in Ref.~\cite{anddd} 

\subsection{Carrier-type classification}

Our first step to re-define the carrier types is to start with the well established semiconductors such as donor- and acceptor-doped Si, prior to analyzing $\alpha$-Fe$_2$O$_3$ and related oxides. The IET requires us to determine the energy-level spacing between the dopant (acceptor or donor) and the host (matrix) material, Si. Undoped Si is an intrinsic semiconductor, in which, the carrier-type can be regarded as an $n$-type due to the dominant conductivity contribution of excited electrons, while the conductivity due to holes strictly require the contribution of the strongly bound electrons in the valence band. Therefore, $n$-type conductivity is relatively larger than the $p$-type for intrinsic semiconductors, including for defect-free $\alpha$-Fe$_2$O$_3$. Furthermore, higher temperatures may also favor $n$-type as opposed to $p$-type due to high conductivity of excited electrons. Note here that we have invoked the principle of least action that determines the path and/or the carrier type with least resistance. From now onwards, we will drop the term $\alpha$ attached to iron oxides for convenience.

Now, we can start consider the extrinsic Si based $n$- and $p$-type semiconductors. For example, the doped element As in Si$^{4+}_{1-x}$As$^{5+}_x$ is a donor with charge $+e$ that attracts a free charge $-e$ from the surrounding host atoms. Moreover, the binding energy of an electron ($-e$) to the $+e$ (As ion center) is assumed to be small compared to the band gap ($E_g$) that introduce a donor impurity level just below (due to small binding energy) the conduction band. If the dopant is B as in Si$^{4+}_{1-x}$B$^{3+}_x$, then B acts as an acceptor with charge $-e$ that attracts the free holes from the host Si material. The acceptor impurity level is now just above the valence band, again due to small binding energy of a hole to a $-e$ (B acceptor center). Obviously, the donor impurities attract the electrons from the host Si, which makes them to be $n$-type. Whereas, acceptors attract holes from the surrounding Si matrix, which in turn one obtains the $p$-type carriers. 

However, it is important to mention here that there are three implicit assumptions for the above-stated explanations$-$ (i) the binding energy of an electron (due to donor) or hole (due to acceptor) should be small relative to $E_g$, (ii) the valence state of an acceptor must be less than the host material, and the opposite is true for a donor (this is physically necessary), and (iii) the existence of the conduction band (semiclassical picture). It should be clear here that conditions (i) and (ii) cannot be defined rigorously for oxides such as Fe$_2$O$_3$, as well as cannot be used directly to explain the $n$- to $p$-type transition. Therefore, we will present a refined, a more general and rigorous classification for the carrier types based on the IET formalism. 

\begin{figure}
\scalebox{0.4}{\includegraphics{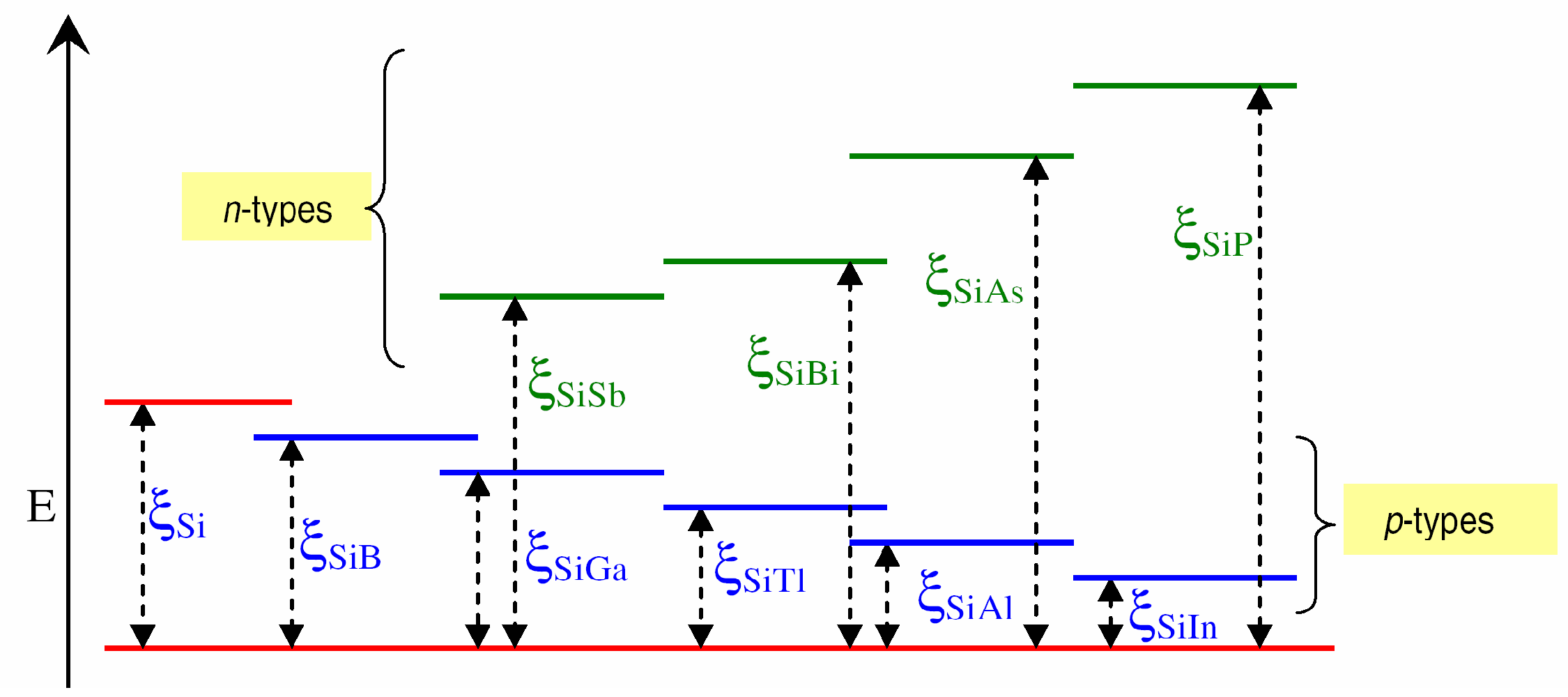}}
\caption{Generalized classification for the well-known $n$- and $p$-type semiconductors. The $p$-types satisfy condition $\mathcal{A}1$, while the $n$-types follow condition $\mathcal{B}2$. The ionization energies of the acceptors, B, Ga, Tl, Al and In can be ordered as $\xi_{\rm Si}^{4+} > \xi_{\rm B}^{3+} > \xi_{\rm Ga}^{3+} > \xi_{\rm Tl}^{3+} > \xi_{\rm Al}^{3+} > \xi_{\rm In}^{3+}$, while, for donors we have $\xi_{\rm Si}^{4+} < \xi_{\rm Sb}^{5+} < \xi_{\rm Bi}^{5+} < \xi_{\rm As}^{5+} < \xi_{\rm P}^{5+}$. Hence, we can understand the systematic changes to the energy-level spacing for both $n$- and $p$-type semiconductors shown in the figure. Note that the energy levels are not to scale. See text for details.}
\label{fig:1}
\end{figure}

The new classification (I) does not require assumption (iii), (II) assumption (ii) is only required for $p$-types due to the principle of least action given earlier, (III) assumption (i) can be replaced by the energy-level spacing and (IV) an additional condition on the elemental composition needs to be invoked for mathematical rigor. The motivation to generalize the original carrier-type classification given in Ref.~\cite{ash} is to explain the reversible $n$- to $p$-type transition in materials other than the traditional ones listed below. Figure~\ref{fig:1} shows all the energy-level spacing and the IET classification for well-established $n$- and $p$-type semiconductors. In particular, we have correctly classified Si$^{4+}$Sb$^{5+}$, Si$^{4+}$Bi$^{5+}$, Si$^{4+}$As$^{5+}$ and Si$^{4+}$P$^{5+}$ as the $n$-types, whereas the $p$-types are given by Si$^{4+}$B$^{3+}$, Si$^{4+}$Ga$^{3+}$, Si$^{4+}$Tl$^{3+}$, Si$^{4+}$Al$^{3+}$ and Si$^{4+}$In$^{3+}$ in accordance with Ref.~\cite{ash} 

The averaged atomic ionization energies for all the elements considered here are given in Table~\ref{Table:I}, in which, the averaging follows Ref.~\cite{and1} Prior to averaging, all the atomic ionization energies were taken from Ref.~\cite{web} From Fig.~\ref{fig:1}, $p$- and $n$-type semiconductors have been reclassified based on these two respective conditions,

\begin {eqnarray}
&&\rm{condition~ \mathcal{A}1}:~~\xi_{\texttt{A}}^{\textit{a}+} < \xi_{\texttt{H}}^{\textit{h}+}\wedge\textit{a} < \textit{h}\wedge\textit{x}_{\texttt{A}} < \textit{y}_{\texttt{H}}. \nonumber \\&& 
\rm{condition~ \mathcal{B}2}:~~\neg \mathcal{A}1, \nonumber
\end {eqnarray}

where $\xi_{\texttt{A,D}}^{a+,d+}$ denotes the acceptors or donors averaged ionization energy, with $a$ and $d$ being their respective valence states. Furthermore, $y_{\texttt{H}} = 1-x_{\texttt{A,D}}$, $y_{\texttt{H}}$ is the host (matrix) element composition and $x_{\texttt{A,D}}$ is the acceptor or donor composition, for example, the elemental composition for Si-B and Si-As systems can be written as Si$^{4+}_{y_{\texttt{H}}}$B$^{3+}_{x_{\texttt{A}}}$ and Si$^{4+}_{y_{\texttt{H}}}$As$^{5+}_{x_{\texttt{D}}}$, respectively, and $\{\textit{a},\textit{h},\textit{d}\} \in \mathbb{N}$, where $\mathbb{N}$ is the set of natural numbers including zero. The symbol $\wedge$ implies that $\xi_{\texttt{A}}^{\textit{a}+} < \xi_{\texttt{H}}^{\textit{h}+}$ \textit{and} $\textit{a} < \textit{h}$ \textit{and} $\textit{x}_{\texttt{A}} < \textit{y}_{\texttt{H}}$ must be satisfied simultaneously. The $p$-type classification given in condition $\mathcal{A}1$ means that the ionization energy of the dopant (acceptor) \textit{and} its valence state \textit{and} its elemental content must be less than the host material (Si$^{4+}$ in this case). Whereas, $\neg \mathcal{A}1$ means any other condition that is \textit{not} $\mathcal{A}1$. Condition $\mathcal{A}1$ can be violated in seven different ways, in which six of the violations point toward the $n$-type carriers, while the last one is equivalent to $\mathcal{A}1$ as proven in the last section (Mathematical analysis).

\begin{table}[ht]
\caption{Averaged atomic ionization energies ($\xi$) for individual ions, and the averaging were carried out based on the valence state for each element. These elements are arranged with increasing atomic number $Z$ for clarity. The unit kJmol$^{-1}$ is adopted for numerical convenience.} 
\begin{tabular}{l c c c } 
\hline\hline 
\multicolumn{1}{l}{Ion}            &    ~~~Atomic number  & ~~~Valence    & ~~~$\xi$   \\  
\multicolumn{1}{l}{}                &   ~~~$Z$             & ~~~state      & ~~~(kJmol$^{-1}$)\\  
\hline 

B                                   &  5	   	  			  &  3+      & 2296 \\ 
C																		&	 6								&	 2+	     & 1720 \\
O																		&  8                &  2+		   & 2351 \\
Al                                  &  13  					    &  3+      & 1713 \\ 
Si	                                &  14 					    &  4+      & 2488 \\
P	                                  &  15 					    &  5+      & 3414 \\
Fe                                  &  26 					    &  2+      & 1162 \\ 
Fe                                  &  26 					    &  3+      & 1760 \\ 
Ga                                  &  31 					    &  3+      & 1840 \\ 
Ge                                  &  32 					    &  4+      & 2503 \\ 
As                                  &  33 					    &  5+      & 3272 \\ 
In                                  &  49 					    &  3+      & 1694 \\
Sn                                  &  50 					    &  4+      & 2248 \\  
Sb                                  &  51 					    &  5+      & 2906 \\ 
Tl                                  &  81 					    &  3+      & 1813 \\ 
Bi                                  &  83 					    &  5+      & 2910 \\ 
 
\hline  
\end{tabular}
\label{Table:I} 
\end{table}


Subsequently, one can write down the mechanism for the $p$-type conductivity as follows: lower ionization energy of the acceptor dopant (\texttt{A}$^{a+}$), compared to its host material implies easier electronic polarizations and/or excitations of the electrons from the acceptor ion centers toward the holes at Si$^{4+}$ ion sites. The required relationship between ionization energy and the polarizability is given in Eq.~(\ref{eq:2}). The electron-polarization from the acceptor to the host ions implies the attraction of holes from the host to the acceptors, which gives rise to $p$-type conductivity. This mechanism is based on condition $\mathcal{A}1$ that is also in agreement with the original mechanism explained in Ref.~\cite{ash} Further increasing the acceptor content (Si$^{4+}_{1-x}$\texttt{A}$^{3+}_x$) reduces the energy-level spacing and therefore, increasing the probability of excitations of the holes from the Si host material.

On the other hand, to obtain $n$-type carriers, the donor dopant, $\texttt{D}^{d+}$ in Si$^{4+}_{1-x}$\texttt{D}$^{d+}_x$ needs to satisfy a somewhat less restrictive sub-conditions, which are derived from the primary condition $\mathcal{B}2$ (see the section on mathematical analysis for details).    

\section{Reversible carrier-type transition in oxides} 

Having understood the reason why and how the polarizability of cations with different valence states and ionization energies affect the Si based semiconductors, we can now extend this knowledge to evaluate the reversible carrier-type transition in Fe$_2$O$_3$. Recall here that the diffusion of oxygen ions creates oxygen vacancies, which in turn have given rise to Fe$^{3+}$ $\rightarrow$ Fe$^{2+}$. The reversible transition is examined by changing the energy-level spacing based on the constituent atoms composition in a given oxide. 


\begin{figure}
\scalebox{0.35}{\includegraphics{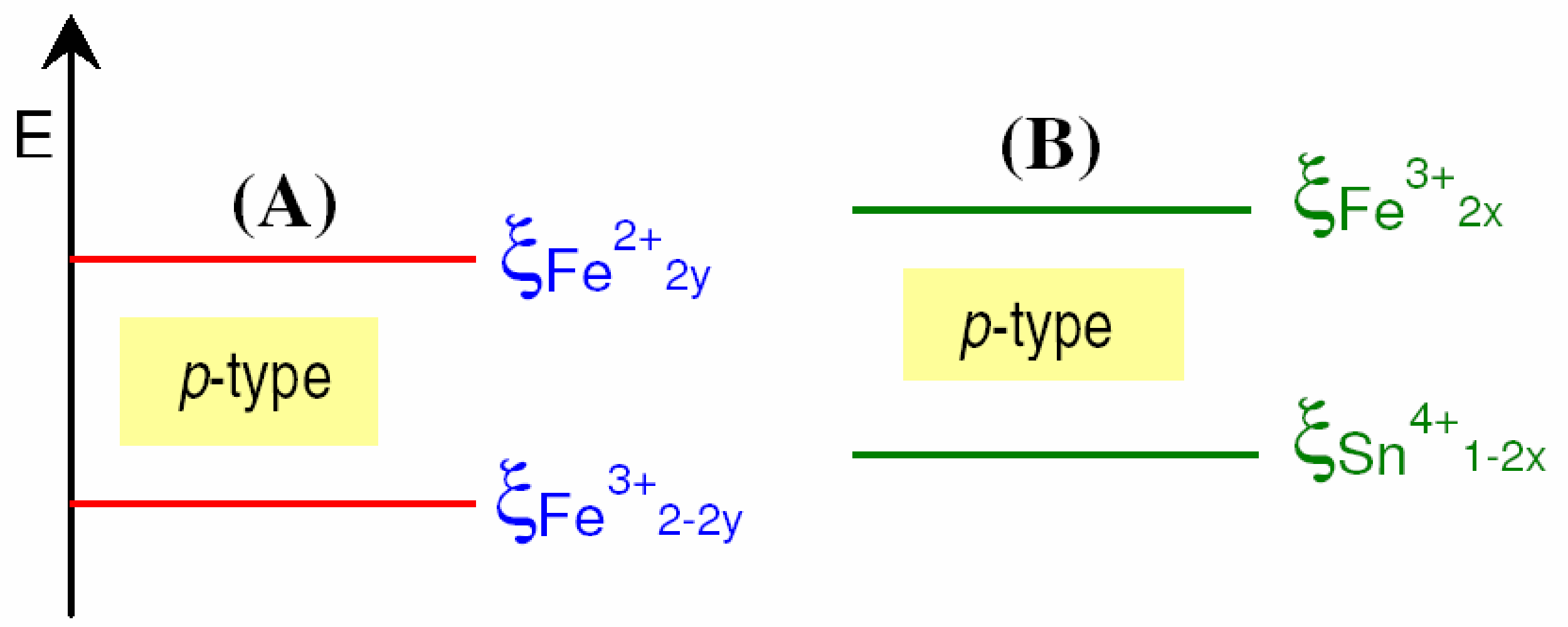}}
\caption{(A) Based on condition $\mathcal{A}1$, $\texttt{A}$ = Fe$^{2+}$, $\texttt{H}$ = Fe$^{3+}$, $\xi_{\rm Fe^{2+}} < \xi_{\rm Fe^{3+}}$, and $0 < y < \frac{1}{2}$ (small Fe$^{2+}$ content), and these properties strictly imply Fe$^{2+}_{2y}$Fe$^{3+}_{2-2y}$O$^{2-}_{3-y}$ is a $p$-type semiconductor for $y \neq 0$. (B) Similarly, Sn$^{4+}_{1-x}$Fe$^{3+}_{x}$O$^{2-}_{2-x}$ is also a $p$-type semiconductor by identifying $\texttt{A}$ = Fe$^{3+}$, $\texttt{H}$ = Sn$^{4+}$, $\xi_{\rm Fe^{3+}} < \xi_{\rm Sn^{4+}}$, $x <\frac{1}{4}$ and $x \neq 0$. Their respective values for $\xi$ are given in Table~\ref{Table:I}. The energy levels are not to scale and the oxygen content are different in both systems. Therefore, one cannot compare system (A) with (B) quantitatively, except for $2 - x = 3 - y$. See text for details.}
\label{fig:2}
\end{figure}

Figure~\ref{fig:2} shows the energy level spacing, and $n$- to $p$-type transition and classifications with respect to condition $\mathcal{A}1$ for Fe$^{2+}_{2y}$Fe$^{3+}_{2-2y}$O$^{2-}_{3-y}$ and Sn$^{4+}_{1-2x}$Fe$^{3+}_{2x}$O$_{2-x}^{2-}$. Here, $y$ is bounded in $\{y:y \leq 1\} \in \mathbb{R}$ while $x$ is bounded in $\{x:x\leq\frac{1}{2}\} \in \mathbb{R}$, where $\mathbb{R}$ is the set of real numbers. These bounds mean that if $y = 1$, then 

\begin {eqnarray}
\rm Fe^{2+}_{2\textit{y}}Fe^{3+}_{2-2\textit{y}}O^{2-}_{3-\textit{y}} ~\rightarrow ~Fe^{2+}O^{2-}, \nonumber
\end {eqnarray}

and if $y = 0$, then 

\begin {eqnarray}
\rm Fe^{2+}_{2\textit{y}}Fe^{3+}_{2-2\textit{y}}O^{2-}_{3-\textit{y}} ~\rightarrow ~Fe^{3+}_{2}O^{2-}_{3}. \nonumber
\end {eqnarray}

Similarly, if $x$ = $\frac{1}{2}$, 

\begin {eqnarray}
\rm Sn^{4+}_{1-2\textit{x}}Fe^{3+}_{2\textit{x}}O_{2-\textit{x}}^{2-} ~\rightarrow ~Fe_2^{3+}O_{3}^{2-}, \nonumber 
\end {eqnarray}

and if $x = 0$, 

\begin {eqnarray}
\rm Sn^{4+}_{1-2\textit{x}}Fe^{3+}_{2\textit{x}}O_{2-\textit{x}}^{2-} ~\rightarrow ~Sn^{4+}O_{2}^{2-}. \nonumber
\end {eqnarray}

On the one hand, substitution of Fe$^{3+}$ into Sn$^{4+}$ sites creates oxygen vacancies, O$_{2-x}$ as a result of different valence states between Sn$^{4+}$ and Fe$^{3+}$. On the other hand, oxygen vacancies ($y$) initiates the transition, Fe$^{3+}$ $\rightarrow$ Fe$^{2+}$ for the oxide Fe$^{3+}_{2}$O$^{2-}_{3}$. In all these cases, the ionization energy approximation can be applied with respect to different cation content ($y$ or $x$).   

\subsection{Iron oxides: Fe$_2$O$_3$, FeO and Fe$_2$O} 

Figure~\ref{fig:2}(A) indicates the energy levels for Fe$^{2+}$ and Fe$^{3+}$ where $\xi_{\rm Fe^{2+}}$ $<$ $\xi_{\rm Fe^{3+}}$ from Table~\ref{Table:I}. We now invoke condition $\mathcal{A}1$ with which, one can associate the acceptor $\texttt{A}$ = Fe$^{2+}$, the host $\texttt{H}$ = Fe$^{3+}$, and therefore, $\xi_{\texttt{A}^{a+}} < \xi_{\texttt{H}^{h+}}$ where $a < h$. In addition, $\textit{x}_{\texttt{A}} < \textit{y}_{\texttt{H}}$ $\Leftrightarrow$ 2$y$ $<$ $2-2y$ due to the fact that $0 < y < \frac{1}{2}$. The reason $y$ should be less than $\frac{1}{2}$ is solely to invoke the existence of small Fe$^{2+}$ concentration as a result of small concentration of oxygen vacancies. In the absence of vacancies, $y = 0$ and all iron ions exist as Fe$^{3+}$. On the other end with $y = 1$, only Fe$^{2+}$ ions exist. As a consequence, Fe$^{2+}_{2y}$Fe$^{3+}_{2-2y}$O$^{2-}_{3-y}$ is a bulk $p$-type semiconductor, and it is a robust one because $n$-type can only be obtained if $y = 0$ (defect-free Fe$_{2}$O$_{3}$) or $y = 1$ (defect-free FeO). 

However, if one were to create interstitial oxygen defects in FeO in such a way to obtain Fe$^{3+}_{2z}$Fe$^{2+}_{1-2z}$O$^{2-}_{1+z}$, we will end up with $n$-type semiconductor. In this oxygen-interstitial case, Fe$^{3+}_{2z}$Fe$^{2+}_{1-2z}$O$^{2-}_{1+z}$ will satisfy one of the seven conditions ($\mathcal{B}_32$) that violates $\mathcal{A}1$ (see the section on mathematical analysis for details). Here, we first need to recognize Fe$^{3+}$ as the donor, $\texttt{D}$, Fe$^{2+}$ as the host, $\texttt{H}$, and therefore $\xi_{\texttt{D}^{d+}} > \xi_{\texttt{H}^{h+}}$ where $d > h$. Subsequently, we can write $\textit{x}_{\texttt{D}} < \textit{y}_{\texttt{H}}$ $\Leftrightarrow$ 2$z$ $<$ $1-2z$ due to $0 < z < \frac{1}{4}$ where $z$ is bounded in $0 \leq z \leq \frac{1}{2}$. If $z = 0$, Fe$^{3+}_{2z}$Fe$^{2+}_{1-2z}$O$^{2-}_{1+z}$ $\Rightarrow$ FeO, whereas, Fe$^{3+}_{2z}$Fe$^{2+}_{1-2z}$O$^{2-}_{1+z}$ $\Rightarrow$ Fe$_2$O$_3$ for $z = \frac{1}{2}$. 

If we introduce oxygen vacancies into FeO, then one obtains Fe$^{+}_{2v}$Fe$^{2+}_{1-2v}$O$^{2-}_{1-v}$. Subsequently, we identify $\texttt{A}$ as Fe$^{+}$ and $\texttt{H}$ as Fe$^{2+}$. Hence, $\xi_{\texttt{A}^{a+}} < \xi_{\texttt{H}^{h+}}$ where $a < h$. As such, we can write $\textit{x}_{\texttt{A}} < \textit{y}_{\texttt{H}}$ $\Leftrightarrow$ 2$v$ $<$ $1-2v$ due to $0 < v < \frac{1}{4}$ where $v$ is bounded in $0 \leq v \leq \frac{1}{2}$. These identifications and conclusions clearly satisfy condition $\mathcal{A}1$, which indicate that Fe$^{+}_{2v}$Fe$^{2+}_{1-2v}$O$^{2-}_{1-v}$ is a $p$-type semiconductor. Finally, if $v = 0$, we obtain FeO and Fe$_2$O for $v = \frac{1}{2}$ from Fe$^{3+}_{2v}$Fe$^{2+}_{1-2v}$O$^{2-}_{1-v}$.

\subsection{Tin oxides: Fe doped SnO$_2$}

Having explained the $n$- and $p$-type classifications for iron oxides, we can now use the similar arguments for Fe doped tin oxides, Sn$^{4+}_{1-2\textit{x}}$Fe$^{3+}_{2\textit{x}}$O$_{2-\textit{x}}^{2-}$. Figure~\ref{fig:2}(B) plots the energy levels for Fe$^{3+}$ and Sn$^{4+}$ that satisfy $\xi_{\rm Fe^{3+}} < \xi_{\rm Sn^{4+}}$. We can recognize that $\texttt{H}^{h+}$ = Sn$^{4+}$, $\texttt{A}^{a+}$ = Fe$^{3+}$ where $a < h$. The presence of Fe$^{3+}$ initiates oxygen vacancies and since the concentration of Fe$^{3+}$ is small, we have x$_{\texttt{A}}$ $<$ y$_{\texttt{H}}$ $\Leftrightarrow$ $2x < 1-2x$ and all these inequalities satisfy condition $\mathcal{A}1$. Therefore, Sn$^{4+}_{1-2\textit{x}}$Fe$^{3+}_{2\textit{x}}$O$_{2-\textit{x}}^{2-}$ is a $p$-type semiconductor within these well-defined inequalities provided that $0 < x < \frac{1}{4}$. However, $x$ is bounded in $0 \leq x \leq \frac{1}{2}$ with which, one obtains SnO$_{2}$ for $x = 0$, and Fe$_2$O$_3$ for $x = \frac{1}{2}$.     

Unfortunately, we are unable to directly compare system (A), Fe$^{2+}_{2y}$Fe$^{3+}_{2-2y}$O$^{2-}_{3-y}$ with system (B), Sn$^{4+}_{1-2\textit{x}}$Fe$^{3+}_{2\textit{x}}$O$_{2-\textit{x}}^{2-}$ where the energy levels and the spacing in Fig.~\ref{fig:2} are for Fe$^{2+}$, Fe$^{3+}$ and Sn$^{4+}$. This means that we did not take oxygen concentration into account because oxygen is an anion. In our discussion earlier, the comparison based on cations are valid because we were only interested in the carrier-type transition that heavily relies on the cations valence states. Nevertheless, we can compare these two systems provided that their [system (A) and (B)] oxygen content are the same, i.e., $2 - x = 3 - y$. This latter condition implies that the energy-level spacing for Fe$_2$O$_2$ (= FeO) $<$ SnO$_2$ (due to $\xi_{\rm Fe^{2+}} < \xi_{\rm Sn^{4+}}$) which is in accordance with the band gap measurements given in Refs.~\cite{band,band2} The optical $E_g$ values for FeO and SnO$_2$ are 2.4 and 3.9 eV, respectively.  

\section{Reversible carrier-type transition during gas-sensing operation} 

In the previous sections, we have explained why and how oxygen defects and cation-doping influenced the changes to the cation valence states, which in turn initiated the carrier-type transition on the oxide surfaces [due to $\textsc{X1}$ only]. Here, we will study the effect of the changes to the chemical compositions of the adsorbed species (or its concentrations) [due to $\textsc{Y1}$: these species can be atoms, ions or molecules] \textit{in the presence} of $\textsc{X1}$. It is important to note here that one need to first identify the carrier-type of oxide surfaces, before interpreting the carrier-type transition due to physisorption. The situation is slightly complicated for nanowires because surface conductivity is only applicable for thin or thick solid films, whereas, for solid nanowires, the conduction is through the whole nanowires, and therefore defect densities strongly affect the conductivity. For example, the conduction mechanism along a single-wall carbon nanotube is indeed complicated when one has to deal with defects along the conducting path.~\cite{marco} 

In any case, we have already explained the reasons why and how a given gas molecule intrinsically prefers a particular oxide surface for optimum sensitivity as a result of different polarizabilities between the gas molecule and the oxide surface.~\cite{oxide} The carrier-type transition model based on the IET formalism presented here does not take into account the gas molecules reacting on the surfaces of the oxide materials to create oxygen vacancies. Hence, the oxygen-vacancy model described in Ref.~\cite{gurlo} will be suitable to study the surface chemical reactions, and then followed by the IET analysis to evaluate the carrier-type transition.   

The following section develops the analysis required for $n \rightarrow p \rightarrow n$ transitions in SnO$_2$ and Fe-doped SnO$_2$ semiconductors in the presence of O$_2$ molecules. Subsequently, we explain how and why the said transitions are influenced by both $\textsc{X1}$ and $\textsc{Y1}$. Next, we apply the IET based carrier-type classification to understand the same transitions occurring in the $\alpha$-Fe$_2$O$_3$ nanowires in the presence of reductive ambient. Unlike SnO$_2$ samples stated above, we will find that, only $\textsc{X1}$ is responsible for the $n \rightarrow p \rightarrow n$ transitions in $\alpha$-Fe$_2$O$_3$ nanowires. Finally, the bulk $\alpha$-Fe$_2$O$_3$ samples that are exposed to O$_2$ and followed by CO molecules give rise to the $n \rightarrow p \rightarrow n$ transitions entirely due to $\textsc{Y1}$. Hence, we have three types of different samples that have different causes (due to $\textsc{X1}$ and/or $\textsc{Y1}$) for carrier-type transitions. However, all of these causes will be shown to have a single origin, which is due to the difference in the polarizability of electrons from the oxide surface and the physisorbed molecules.

\subsection{SnO$_2$ and Fe-doped SnO$_2$ in the presence of O$_2$ molecules} 

We first recall the results reported by Galatsis et al.~\cite{gala}, in which, they found that a pure SnO$_2$ is an $n$-type semiconductor, and after doping with Fe, it started to behave like a $p$-type semiconductor when both doped and undoped samples are exposed to O$_2$ molecules.~\cite{gala} This $n$- to $p$-type transition is in accordance with our analysis given earlier where at a lower temperature, there is a lower probability for electrons to be excited to conduct electricity. In addition, we assume here that O$_2$ molecules attract electrons from the oxide surface (smaller polarizability of O$_2$ molecule compared to oxide surfaces), which will enhance the concentration of surface holes, and therefore, Fe doped SnO$_2$ is expected to remain as a $p$-type semiconductor for a lower O$_2$ concentration and at a lower temperature (350$^{\rm o}$C). This assumption is valid if the oxide surfaces contain dangling bonds and low valence state defects (high valence state defects will reduce the polarizability of the oxide surface). However, note here that $p$- to $n$-type transition is possible at a much higher O$_2$ concentration. Consequently, we can surmise here that $n$-type conductivity is favored at high temperatures, while higher oxygen concentration favors $p$-type carriers (due to large electron affinity or small polarizability of O$_2$ molecules). This means that adsorbed oxygen species competes with increasing temperatures to maintain $p$-type conductivity. 

\begin{figure}
\scalebox{0.2}{\includegraphics{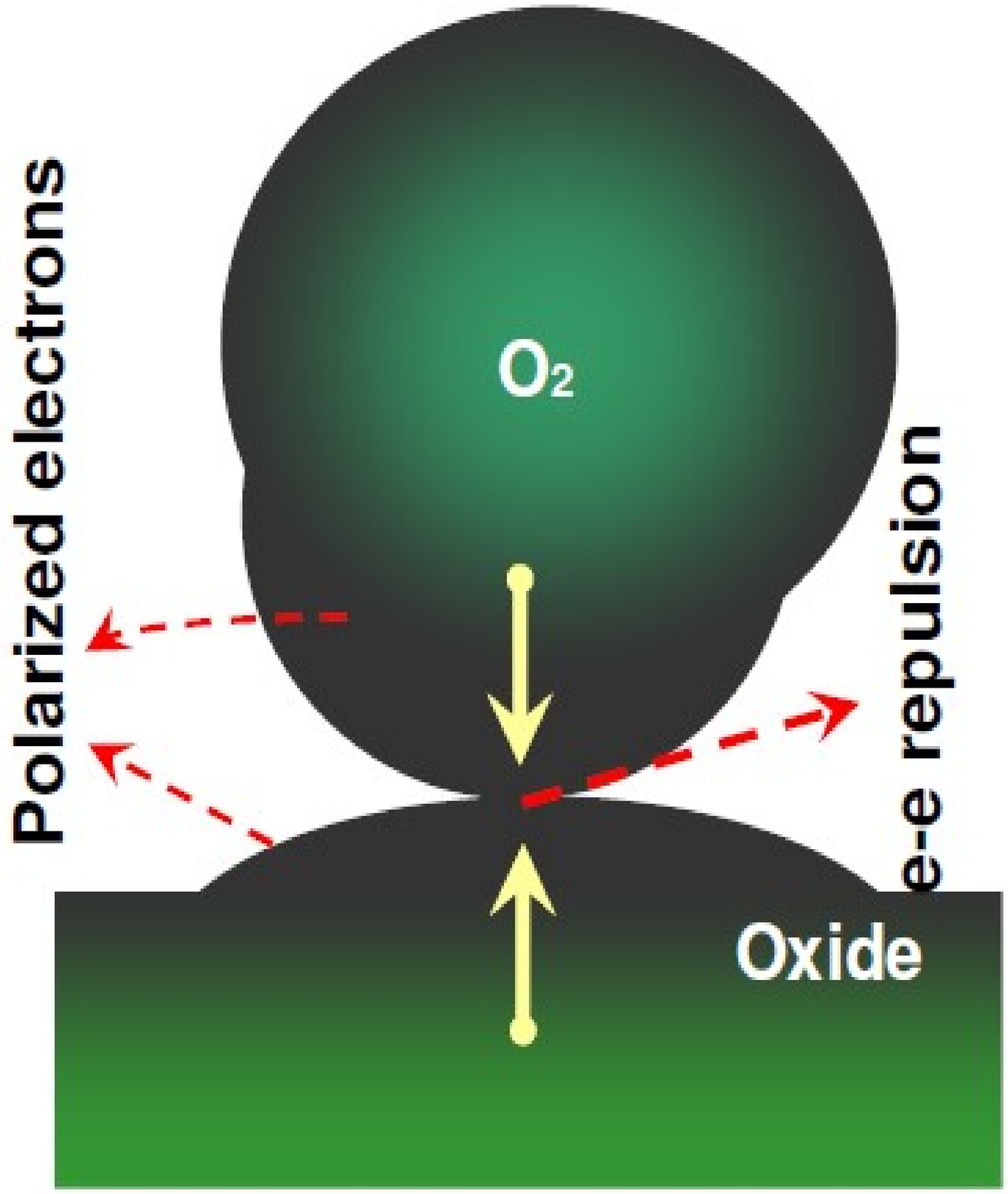}}
\caption{The larger polarizability magnitude for both the O$_2$ molecule and the oxide surface implies these two strongly polarized entities may repel each other, giving rise to an effective dissociation and negligible conductance response. The magnitude of the electron-electron ($e$-$e$) repulsion, size of the molecules and the electron polarization are not to scale. See text for details.}
\label{fig:3}
\end{figure}

Having said that, we can now explore the experimental data measured at 400$^{\rm o}$C, where $p$- to $n$-type transition has been observed for 0.1 and 1\% of O$_2$ concentrations.~\cite{gala} This transition is entirely due to temperature effect where O$_2$ content is insufficient to maintain $p$-type conductivity. As anticipated, much higher concentration of O$_2$ (10\%) have attracted (due to low polarizability) the excited electrons (due to 400$^{\rm o}$C $>$ 350$^{\rm o}$C) from the oxide surface so as to maintain the $p$-type conductivity. Finally, for 450$^{\rm o}$C and 10\% of O$_2$, $p$- to $n$-type transition was observed, which means that 10\% of O$_2$ molecules are insufficient to attract the additional excited electrons due to a much higher temperature (450$^{\rm o}$C $>$ 400$^{\rm o}$C) from the oxide surface. However, a reverse $n$- to $p$-type transition may not be observable for a much higher O$_2$ content ($>$ 10\%) at 450$^{\rm o}$C due to polarized electron-electron repulsion from both the polarized oxide surface and O$_2$ molecules~\cite{oxide} (see Fig.~\ref{fig:3}). In addition, at a much higher temperature ($>$ 450$^{\rm o}$C), chemical reactions between O$_2$ and a given oxide surface could be the dominant surface process with O atoms diffuse into the oxide surface. Here, we have explained why both $\textsc{X1}$ (due to Fe doping) and $\textsc{Y1}$ (due to adsorbed O$_2$ molecules) are responsible for the carrier-type transitions during gas-sensing operations.  


Finally, SnO$_2$ has remained as an $n$-type throughout the measurements for different O$_2$ concentrations (0.1 to 10\%) and temperatures (350 to 450$^{\rm o}$C).~\cite{gala} This is simply because SnO$_2$ does not satisfy condition $\mathcal{A}1$, in which, there are no lower concentration of acceptor ions with smaller valence states (compared to Sn$^{4+}$), and with smaller ionization energy to create holes as charge carriers. 

\subsection{$\alpha$-Fe$_2$O$_3$ nanowires in the presence of reductive ambient}

We now recall the results of Lee et al.~\cite{lee} They have made $p$-type $\alpha$-Fe$_2$O$_3$ nanowire semiconductor by creating oxygen vacancies inside the nanowire, rather than on the surface of the nanowire. This conclusion is supported by their EELS measurements, which confirms the nanowire surfaces contain negligible amount of oxygen vacancies, while inside the $\alpha$-Fe$_2$O$_3$ nanowire one finds highly ordered oxygen vacancies.~\cite{lee} From our analysis earlier, $\alpha$-Fe$_2$O$_3$ with oxygen vacancies should be a $p$-type semiconductor (due to the existence of Fe$^{2+}$ in the presence of oxygen vacancies). This $p$-type conductivity is observable because the conduction is through the whole nanowire instead of only on the nanowire surface. The reason is that if the conduction was to occur only on the nanowire surface (that contains insufficient oxygen vacancies), then one should observe $n$-type conductivity based on the IET analysis and EELS measurements. Interestingly, our analysis are in complete agreement with the experimental results and proposals reported in Ref.~\cite{lee}, where oxygen-vacancy ordering in the centre is responsible for the $p$-type conductivity. 

\begin{figure}
\scalebox{0.4}{\includegraphics{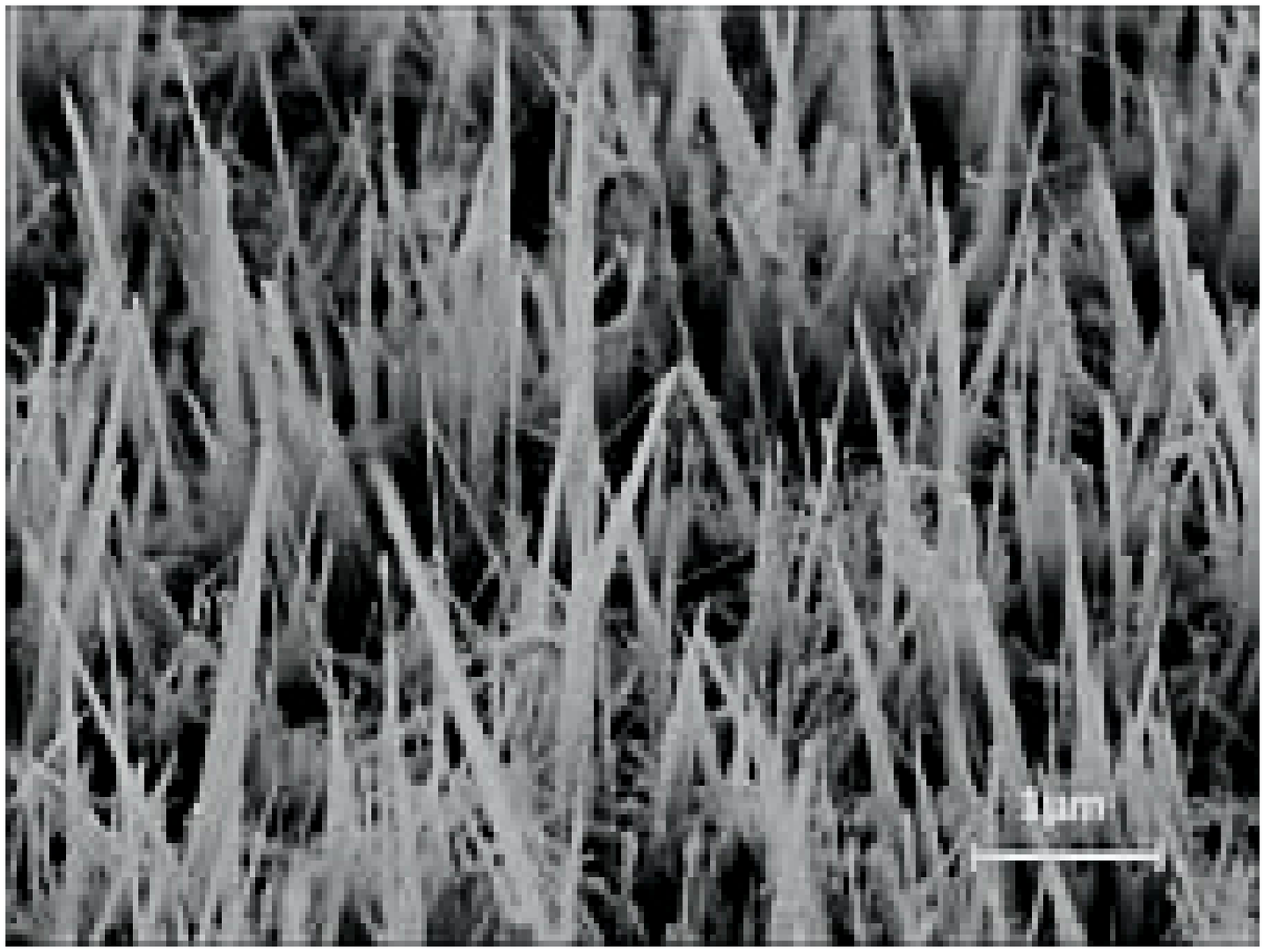}}
\caption{$\alpha$-Fe$_2$O$_3$ nanowires with oxygen-vacancy ordering synthesized on an iron metal surface after direct low temperature plasma oxidation in radio-frequency (RF) oxygen. Visible oxygen vacancies in these nanowires have been captured via the TEM images, which are given in Refs.~\cite{uros2,chen}.}
\label{fig:4}
\end{figure}

Furthermore, exposing this $p$-type $\alpha$-Fe$_2$O$_3$ nanowires to reductive ambient have transformed them into $n$-type nanowires (due to $p$- to $n$-type transition).~\cite{lee} This is interesting given the fact that the EELS measurements~\cite{lee} further confirms the existence of higher oxygen-vacancy densities in the centre, which is also in accordance with previous reports.~\cite{ch,jas} Higher oxygen-vacancy densities imply that the oxygen ordering in the centre has been destroyed via further diffusion of oxygen ions to the surface.~\cite{lee} Parallel to this, we propose here that such a heavy oxygen diffusion process during the reductive ambient annealing will give rise to oxygen accumulation on the nanowire surface with increased surface thickness with stoichiometric $\alpha$-Fe$_2$O$_3$. As a consequence, $n$-type carriers (due to stoichiometric $\alpha$-Fe$_2$O$_3$ on the surface) should dominate the conductivity, in which, high-defect concentration in the centre is no longer favorable for conduction. Again, we have invoked the principle of least action. Recall here that stoichiometric $\alpha$-Fe$_2$O$_3$ means that $\alpha$-Fe$_2$O$_3$ with negligible oxygen vacancies and Fe$^{2+}$. In summary, we have explained the reasons why and how only $\textsc{X1}$ has come to play in determining the carrier-type transitions during gas-sensing operations (due to creation of oxygen-vacancy ordering, and followed by random oxygen vacancies created due to exposure to reductive ambient). Parallel to the thermal method~\cite{lee}, one can also synthesize these $\alpha$-Fe$_2$O$_3$ nanowires using the plasma oxidation method (POM) with a much shorter synthesizing time (120 s) as shown in Fig.~\ref{fig:4}. The nanowires created on the iron-metal surface using the low temperature POM have been highlighted in the introduction, where these nanowires have been found to contain highly ordered oxygen vacancies. The readers are referred to Refs.~\cite{uros2,chen} for more details on these vacancies observed with TEM, and the POM experimental procedure. 


The next sub-section will explain how one can obtain the $n \rightarrow p \rightarrow n$ transitions entirely due to $\textsc{Y1}$, in the absence of $\textsc{X1}$ (the carrier-type on the oxide surfaces are not changed by doping and/or creating oxygen vacancies and/or defects). It is worth to mention here that there is also an earlier transistor-effect study reported on $n$- and $p$-type $\alpha$-Fe$_2$O$_3$ nanobelts.~\cite{fan} In their experiments, Fan et al.~\cite{fan} have used Zn dopant to obtain the carrier-type transition from $n$- to $p$-type by controlling the temperature during the doping process. This transition is due to X1. 

\subsection{$\alpha$-Fe$_2$O$_3$ exposed to O$_2$ and followed by CO molecules}

\begin{figure}
\scalebox{0.4}{\includegraphics{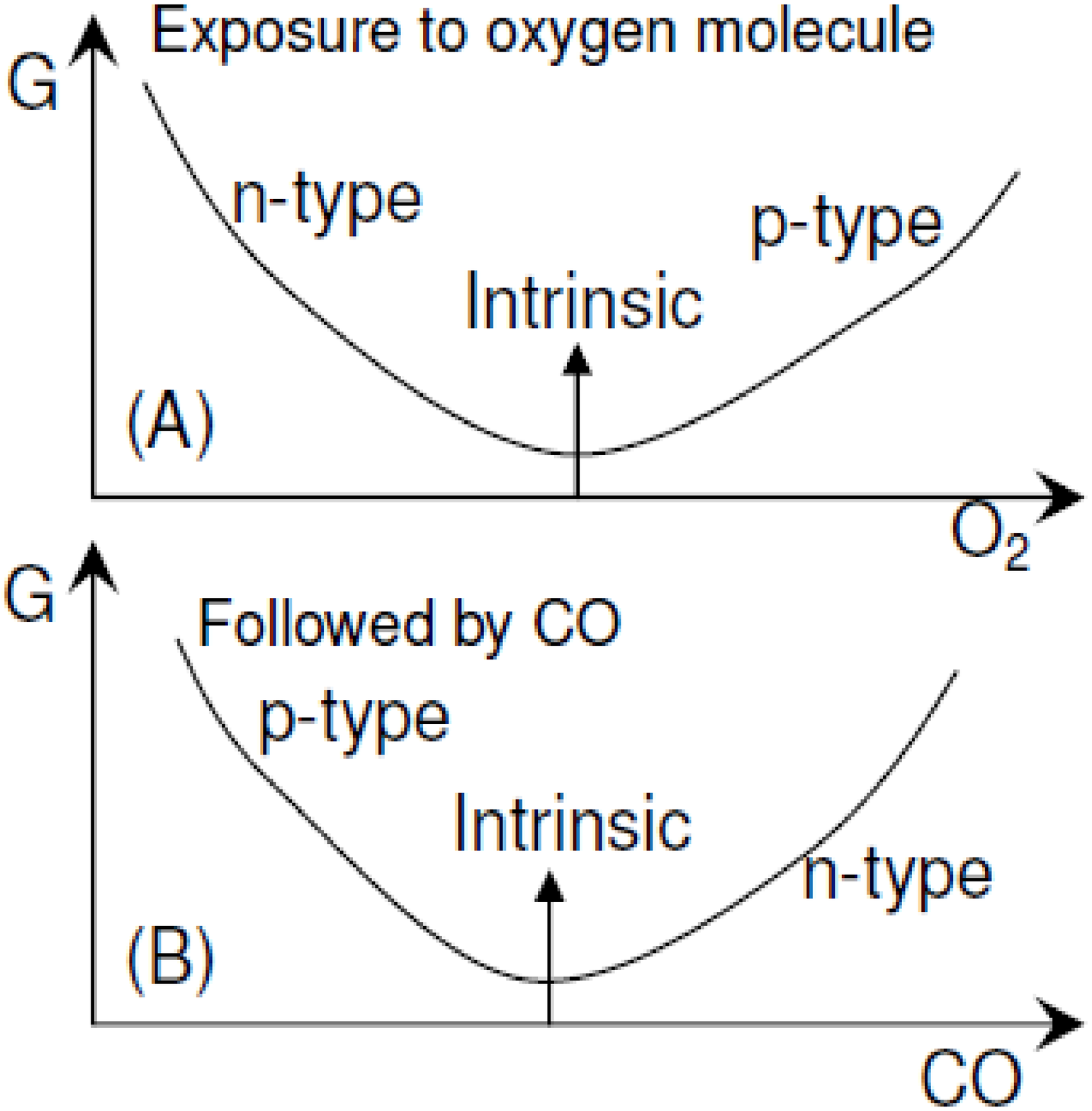}}
\caption{The expected changes to the conductance, G are sketched in (A) due to increasing amount of O$_2$ molecules exposed onto the $\alpha$-Fe$_2$O$_3$ surface. The conductance first decrease due to decreasing electron concentration, and increase after reaching a minimum (intrinsic conductance) as a result of increasing hole concentration. In this case, we obtained the $n$- to $p$-type transition. Subsequently, exposing CO onto the oxygen adsorbed $\alpha$-Fe$_2$O$_3$ surface [as shown in (B)], leads to a reverse transition, namely, $p$- to $n$-type. See text for details.}
\label{fig:5}
\end{figure}

Our aim to understand the carrier-type transition during the gas-sensing operation so far has not directly depended on any assumption on surface energy band bending, work function, and so on as reported by Gurlo et al.~\cite{gur,gurr2} However, we have made three important assumptions along the way to this sub-section, namely, (1) oxygen molecule's polarizability is smaller than the SnO$_2$ and Fe doped SnO$_2$ surfaces, (2) the conducting paths for a nanowire is through the whole nanowire, and not only on its surfaces, and (3) sufficient surface thickness with stoichiometric $\alpha$-Fe$_2$O$_3$ is a preferable conduction path with least resistance compared to the nanowire centres with high and randomly oriented defect densities. Assumption (1) is valid if the surface defects contain (i) cations with smaller valence states such as Fe$^{2+}$ (due to $\xi_{\rm Fe^{2+}} < \xi_{\rm Fe^{3+}}$) and/or (ii) dangling bonds. These dangling bonds are nothing but unpaired or free electrons (not literally free, but bound to the surface parent ion) that did not happen to make any bond due to defects, and usually occur on the surfaces of solids for well-known reasons. Obviously, if the electrons are bonded, then it is much more difficult to polarize them. Both (i) and (ii) will give rise to a larger surface polarizability.~\cite{oxide} Our assumptions in (2) and (3) are also valid because even though charge carriers tend to flow on the surface (skin effect)~\cite{ash}, this is not the case for nanowires with inhomogeneous defect distribution. Unlike the $n \rightarrow p \rightarrow n$ transitions discussed earlier by Galatsis et al.~\cite{gala} and Lee et al~\cite{lee}, which are due to both $\textsc{X1}$ and $\textsc{Y1}$, and $\textsc{X1}$ only, respectively, the same transitions reported by Gurlo et al.~\cite{gur} are entirely due to $\textsc{Y1}$ as pointed out in the introduction. Gurlo et al.~\cite{gur} started out the measurements with an $n$-type $\alpha$-Fe$_2$O$_3$, by first exposing it to O$_2$ gas. In their analysis, sufficient amount of oxygen-ion adsorption on the surfaces (ionosorption model) leads to an inversion layer that changes the conduction-type from $n$- to $p$-type as a result of decreasing conductance (due to decreasing electrons) and followed by increasing conductance (due to increasing holes) with increasing work function (due to increasing oxygen concentration)~\cite{gur} [see Fig.~\ref{fig:5}(A)]. 


Subsequent exposure to CO (reductive ambient), have led CO reacting with adsorbed oxygen ions on the $\alpha$-Fe$_2$O$_3$ surface, releasing electrons from the adsorbed oxygen ions into the surface conduction band. As a consequence, this increasing electron concentration gives rise to smaller magnitude of the surface band bending, in which, with enough concentration of CO, the Fermi level can be brought above the intrinsic level to give $p$- to $n$-type transition [see Fig.~\ref{fig:5}(B)]. In this case, the conductance decreases first due to decreasing hole concentration and with further increasing of CO, the conductance increases again because the concentration of free-electrons is increasing with respect to decreasing band bending.~\cite{gur} Hence, it is clear now that the above stated $n \rightarrow p \rightarrow n$ transitions are indeed because of $\textsc{Y1}$ only. We know from the IET and from our previous discussion (recall the analysis for Fe doped SnO$_2$) O$_2$ molecules have smaller polarizability, which can attract the excited surface electrons (due to temperature), and with increasing O$_2$ concentration one should be able to observe the $n$- to $p$-type transition. In this case, the conductance is expected to be similar to the one observed in Ref.~\cite{gur}, where the conductance first decreases (due to decreasing electrons) and followed by increasing conductance due to increasing holes as a result of increasing O$_2$ concentration (again, this transition is similar to Fe doped SnO$_2$ exposed to O$_2$). On the other hand, introducing CO after O$_2$ exposure is somewhat complicated and our arguments are as follow.

We first need to evaluate the polarizability of CO relative to O$_2$ molecules following the procedure discussed in Ref.~\cite{oxide}. Here, C$^{2+}$ acts as a cation and 2 electrons are transfered from C$^{2+}$ to O$^{2-}$ giving the charge states, C$^{2+}$O$^{2-}$. Such electron-transfers are due to $\alpha_d^{\rm C^{2+}} > \alpha_d^{\rm O^{2+}}$, which is from $\xi_{\rm C^{2+}} < \xi_{\rm O^{2+}}$ (see Table~\ref{Table:I}). 

For an oxygen molecule however, both O atoms in O$_2$ are identically cationic and anionic in electronic structure. This means that there is no effective electron transfer, but we can treat one of the O atom in O$_2$ as a cation, while the other as an anion, and we can switch those O atoms by requiring the former cation as the anion and the previous anion as the cation. Hence, there are 2 electrons contributed by O$^{(1)}$, and another 2 from O$^{(2)}$ atom, and this 4 electrons are responsible for the polarizability of O$_2$ molecule. We can now readily obtain the inequality $\alpha_d^{\rm CO} > \alpha_d^{\rm O_2}$ from $\xi_{\rm C^{2+}} < \xi_{\rm O^{2+}}$. This allows us to conclude that the electrons from O$_2$ molecule is relatively more difficult to be polarized compared to CO molecule. In other words, the assumption (1) made earlier is further supported by the IET analysis, even if the comparison here is with CO, and not with SnO$_2$ or Fe doped SnO$_2$. Add to that, our polarizability evaluation is also strictly valid even though the diatomic bonding energy for CO (1077 kJmol$^{-1}$) is larger than O$_2$ (498 kJmol$^{-1}$) molecule. These bonding energy values were obtained from Ref.~\cite{web}. Smaller bonding energy of a diatomic molecule (such as O$_2$ molecule) that has smaller polarizability relative to a CO molecule means that any attempt to increase O$_2$ polarizability significantly due to temperature will lead to the breaking of the O$-$O bond first, before significant polarizability could occur. Whereas, CO molecule can be easily polarized (compared to O$_2$) without first breaking the C$-$O bond. In fact, we have cautioned the readers of this scenario where one should be careful in interpreting the bonding energies for polarizability calculations.~\cite{and4,and8} Note that the dash sign ``$-$" written above for O$-$O and C$-$O, do not in any way represent the number of bonds.   

Further evidences in support of the IET based polarizability analysis (for O$_2$ and CO) come from the work of Reed~\cite{reed} and Bogaard et al.~\cite{bog} The molecular polarizability for O$_2$ is given by 1.75 $\times$ 10$^{-40}$ Cm$^2$V$^{-1}$ (we have converted the unit from cm$^{3}$ (Gaussian) to SI).~\cite{reed} Whereas, the molecular polarizability~\cite{bog} for CO is 2.031 $\times$ 10$^{-40}$ Cm$^2$V$^{-1}$, which indicate the correctness of our analysis where $\alpha_d^{\rm CO} > \alpha_d^{\rm O_2}$. The conversion coefficient to convert the Gaussian to SI unit reads $4\pi\epsilon_0(10^{-6})$ where $\alpha^{\rm SI}_d$ = $4\pi\epsilon_0(10^{-6})\alpha_d^{\rm Gaussian}$. We stress here that the IET analysis is based on the atomic energy-level spacing renormalization technique~\cite{andrew}, and therefore we did not use the standard variational principle to solve the many-body Hamiltonian. This means that we can derive the required analytic functions for accurate and strictly self-consistent analysis of the physico-chemical mechanisms. Sometimes, these analysis can be made quantitative with remarkable agreements with the experimental results.~\cite{and2}    

We can now revisit the $p$- to $n$-type transition occurring on the oxygen adsorbed $\alpha$-Fe$_2$O$_3$ surface. Contrary to Ref.~\cite{gur}, there is no convincing spectroscopic measurements to justify the existence of oxygen ionosorption, such as O$_2^{-}$, O$^-$ and O$_2^{2-}$.~\cite{gurlo,gurlo2} Parallel to this, we have proposed above that the adsorbed oxygen molecule attract the electrons from the oxide surface and consequently, gives rise to $n$- to $p$-type transition. Subsequent exposure to high concentration of CO molecules reduces the polarizability strength of the adsorbed O$_2$ molecules. The reason is that in the presence of CO, adsorbed O$_2$ molecules will also attract the electrons from the highly polarizable CO molecules (due to $\alpha_d^{\rm CO} > \alpha_d^{\rm O_2}$), thus reducing the O$_2$ strength to attract the electrons from the $\alpha$-Fe$_2$O$_3$ surface. Therefore, we obtain the $p$- to $n$-type transition. 

In summary, Fig.~\ref{fig:5}(A) and (B) are also valid for the IET analysis, in which, our analysis for the SnO$_2$, Fe doped SnO$_2$ and $\alpha$-Fe$_2$O$_3$ samples are all in agreement with the experimental results, as well as highly self-consistent. This completes our analysis and explanations as to why and how carrier-type transition occur during gas-sensing operations. Moreover, the discussion and analysis presented here based on the polarizability effect do not require physical electron transfer from a molecule to the surface or vice versa, which leaves the adsorbed molecules to be charged. The individual system (adsorbed molecules or the oxide surface) during conductivity measurements may remain neutral, but can be strongly polarized to either enhance or reduce the conductance as explained earlier and in Ref.~\cite{oxide}        

\section{Mathematical analysis}

As stated earlier, condition $\mathcal{A}1$ can be violated in seven different ways and the first six will give rise to $n$-type carriers, while the last one is shown to be exactly equal to $\mathcal{A}1$. Here, we list all the possible sub-conditions (seven in total) for condition $\mathcal{B}_k2$ where $k$ = 1, 2, \ldots, 7.  

\begin {eqnarray}
&&\rm{\mathcal{B}_12}:~~\xi_{\texttt{A}}^{\textit{a}+} > \xi_{\texttt{H}}^{\textit{h}+}\wedge\textit{a} < \textit{h}\wedge\textit{x}_{\texttt{A}} < \textit{y}_{\texttt{H}}, \nonumber \\&& 
\rm{\mathcal{B}_22}:~~\xi_{\texttt{A}}^{\textit{a}+} < \xi_{\texttt{H}}^{\textit{h}+}\wedge\textit{a} > \textit{h}\wedge\textit{x}_{\texttt{A}} < \textit{y}_{\texttt{H}}, \nonumber \\&& 
\rm{\mathcal{B}_32}:~~\xi_{\texttt{A}}^{\textit{a}+} < \xi_{\texttt{H}}^{\textit{h}+}\wedge\textit{a} < \textit{h}\wedge\textit{x}_{\texttt{A}} > \textit{y}_{\texttt{H}}, \nonumber \\&& 
\rm{\mathcal{B}_42}:~~\xi_{\texttt{A}}^{\textit{a}+} > \xi_{\texttt{H}}^{\textit{h}+}\wedge\textit{a} > \textit{h}\wedge\textit{x}_{\texttt{A}} < \textit{y}_{\texttt{H}}, \nonumber \\&&
\rm{\mathcal{B}_52}:~~\xi_{\texttt{A}}^{\textit{a}+} > \xi_{\texttt{H}}^{\textit{h}+}\wedge\textit{a} < \textit{h}\wedge\textit{x}_{\texttt{A}} > \textit{y}_{\texttt{H}}, \nonumber \\&&
\rm{\mathcal{B}_62}:~~\xi_{\texttt{A}}^{\textit{a}+} < \xi_{\texttt{H}}^{\textit{h}+}\wedge\textit{a} > \textit{h}\wedge\textit{x}_{\texttt{A}} > \textit{y}_{\texttt{H}}, \nonumber \\&&
\rm{\mathcal{B}_72}:~~\xi_{\texttt{A}}^{\textit{a}+} > \xi_{\texttt{H}}^{\textit{h}+}\wedge\textit{a} > \textit{h}\wedge\textit{x}_{\texttt{A}} > \textit{y}_{\texttt{H}}. \nonumber
\end {eqnarray}

We start with $\mathcal{B}_12$ and construct a general hypothetical system that has the form,  $\texttt{H}_{\textit{y}_{\texttt{H}}}^{h+}\texttt{A}_{\textit{x}_{\texttt{A}}}^{a+}$. Recall here that $\texttt{H}$ and $\texttt{A}$ denote host and acceptor cations, respectively. Next, we invoke the violation, which is given by $\xi_{\texttt{A}} > \xi_{\texttt{H}}$ implying the additional $-e$ at the ${\texttt{A}}^{a+}$ center cannot be easily excited toward the ${\texttt{H}}^{h+}$ centers or conversely, it is difficult for a hole to be attracted toward the ${\texttt{A}}^{a+}$ center. This process favors the electrons from ${\texttt{H}}^{h+}$ as the effective carriers ($n$-type). Note here that $\mathcal{B}_12$ stays intact without any transformation because transformations are only required if $\textit{a} < \textit{h}$ or $\textit{x}_{\texttt{A}} < \textit{y}_{\texttt{H}}$ or both are violated (see below for other sub-conditions). Subsequently, the violation for $\mathcal{B}_22$ is due to $a > h$. In this case, we require the transformation, $\texttt{H}_{\textit{y}_{\texttt{H}}}^{h+}\texttt{A}_{\textit{x}_{\texttt{A}}}^{a+} \rightarrow \texttt{H}_{\textit{y}_{\texttt{H}}}^{h+}\texttt{D}_{\textit{x}_{\texttt{D}}}^{d+}$ due to $a \rightarrow d$ in which, $\texttt{A}_{\textit{x}_{\texttt{A}}}^{a+}$ and $\xi^{a+}_{\texttt{A}}$ have been replaced by $\texttt{D}_{\textit{x}_{\texttt{D}}}^{d+}$ and $\xi^{d+}_{\texttt{D}}$, respectively. This sub-condition still favors electrons entirely due to $\textit{x}_{\texttt{D}} < \textit{y}_{\texttt{H}}$ because if $\textit{x}_{\texttt{D}} > \textit{y}_{\texttt{H}}$, then $d \rightarrow h$ and $h \rightarrow a$ and these latter transformations imply $p$-type. The newly transformed $\mathcal{B}_22$ for $n$-type carriers is given by $\xi_{\texttt{D}}^{\textit{d}+} < \xi_{\texttt{H}}^{\textit{h}+} \wedge \textit{d} > \textit{h} \wedge \textit{x}_{\texttt{D}} < \textit{y}_{\texttt{H}}$. Next, if $\mathcal{B}_32$ is true, then the violations are caused by $\textit{x}_{\texttt{A}} > \textit{y}_{\texttt{H}}$. Consequently, $\texttt{H}_{\textit{y}_{\texttt{H}}}^{h+}\texttt{A}_{\textit{x}_{\texttt{A}}}^{a+} \rightarrow \texttt{H}_{\textit{y}_{\texttt{H}}}^{h+}\texttt{D}_{\textit{x}_{\texttt{D}}}^{d+}$ due to $a \rightarrow h$ and $h \rightarrow d$. Hence, the transformed $\mathcal{B}_32$ is given by $\xi_{\texttt{H}}^{\textit{h}+} < \xi_{\texttt{D}}^{\textit{d}+} \wedge \textit{h} < \textit{d} \wedge \textit{y}_{\texttt{H}} > \textit{x}_{\texttt{D}}$ that gives $n$-type carriers.   

The following three sub-conditions, $\mathcal{B}_42$, $\mathcal{B}_52$ and $\mathcal{B}_62$ deal with any two inequalities that violate condition $\mathcal{A}1$ simultaneously. Invoking $\mathcal{B}_42$ with two violations, $\xi_{\texttt{A}}^{\textit{a}+} > \xi_{\texttt{H}}^{\textit{h}+}$ and $a > h$ impose $a \rightarrow d$ and $\texttt{A}_{\textit{x}_{\texttt{A}}}^{a+} \rightarrow \texttt{D}_{\textit{x}_{\texttt{D}}}^{d+}$ that consequently lead to $\texttt{H}_{\textit{y}_{\texttt{H}}}^{h+}\texttt{A}_{\textit{x}_{\texttt{A}}}^{a+} \rightarrow \texttt{H}_{\textit{y}_{\texttt{H}}}^{h+}\texttt{D}_{\textit{x}_{\texttt{D}}}^{d+}$ and $\xi_{\texttt{D}}^{\textit{d}+} > \xi_{\texttt{H}}^{\textit{h}+} \wedge \textit{d} > \textit{h} \wedge \textit{x}_{\texttt{D}} < \textit{y}_{\texttt{H}}$. Therefore, one can transform $\mathcal{B}_42 \rightarrow \mathcal{B}_32$ to obtain $n$-type carriers. The two violations in sub-condition $\mathcal{B}_52$ are $\xi_{\texttt{A}}^{\textit{a}+} > \xi_{\texttt{H}}^{\textit{h}+}$ and $x^{\textit{a}+} > y_{\texttt{H}}$. We can transform $\mathcal{B}_52$ using $a \rightarrow h$ and $h \rightarrow d$ to obtain $\xi_{\texttt{H}}^{\textit{h}+} > \xi_{\texttt{D}}^{\textit{d}+} \wedge \textit{h} < \textit{d} \wedge \textit{y}_{\texttt{H}} > \textit{x}_{\texttt{D}}$ where $\mathcal{B}_52 \rightarrow \mathcal{B}_22$, which is in agreement with $n$-type carriers. For $\mathcal{B}_62$, $a > h$ and $x_{\texttt{A}} > y_{\texttt{H}}$ give rise to the transformations, $a \rightarrow h$ and $h \rightarrow a$, thus $\mathcal{B}_62$ can be rewritten as $\xi_{\texttt{A}}^{\textit{a}+} > \xi_{\texttt{H}}^{\textit{h}+}\wedge\textit{a} < \textit{h}\wedge\textit{x}_{\texttt{A}} < \textit{y}_{\texttt{H}}$, which is equivalent to $\mathcal{B}_12$ that satisfies $n$-type carriers. Finally, $\mathcal{B}_72$ violates all three inequalities and the required transformations are $a \rightarrow h$ that is due to $x_{\texttt{A}} > y_{\texttt{H}}$ and $h \rightarrow a$ due to $a > h$. After the usual transformations one can obtain $\xi_{\texttt{H}}^{\textit{h}+} > \xi_{\texttt{A}}^{\textit{a}+} \wedge \textit{h} > \textit{a} \wedge \textit{y}_{\texttt{H}} > \textit{x}_{\texttt{A}}$ and this is nothing but $\mathcal{A}1$.

It is not surprising that we have shown the existence of these equalities, $\mathcal{B}_12$ = $\mathcal{B}_62$, $\mathcal{B}_22$ = $\mathcal{B}_52$, $\mathcal{B}_32$ = $\mathcal{B}_42$, and $\mathcal{B}_72$ = $\mathcal{A}1$. This property arises entirely from the symmetry in the mathematical transformation of the inequalities ($> \leftrightarrow <$) in all the sub-conditions stated earlier.   

\section{Conclusions}

We have developed a rigorous and unequivocal strategy to understand the unavoidable oxygen diffusion that creates oxygen vacancies in oxides as a mean to switch the carrier types in iron oxides and Fe doped tin oxides. In addition, we have explained how different gas environments may interact with a given oxide to initiate the reversible carrier-type switching in metal-oxide sensors. This reversible switching can be achieved on the oxide surface entirely due to different gaseous environment or through doping the oxide, and followed by exposure to a particular gas. All these scenarios have been explained self-consistently within the IET based carrier-type classification and polarizability functional. 

We found that the $p$-type semiconducting oxides strictly bounded within a narrow range of certain critical parameters as compared to $n$-type oxides. In particular, $p$-type semiconductors can only be observed if the acceptor's ionization energy ($\xi$), valence state ($a+$) and concentration ($\textit{x}_{\texttt{A}}$) are all less than the host cations. Any other possibility will give rise to $n$-type carriers. However, higher temperatures and/or $x_{\texttt{A}} \ll y_{\texttt{H}}$ may also give rise to $n$-type carriers such that, $p$-type carriers may not be observable. Our refined and extended classification within the IET also agrees with the original mechanism explained for traditional Si based semiconductors. Apart from that, all our predictions can be tested with different oxide surfaces exposed to different gas molecules. The required measurements are the conductance for different temperature and gas concentrations, and the electron energy loss spectroscopy on the oxide surface before and after exposing to different gas environments.    

The knowledge derived from the carrier-type reclassification and its analysis on gas-sensing oxides have provided valuable information for one to exploit and to develop tailor-made breath-sensors (medicine), gas sensors (engineering) and odor sensors (food processing) with specific and improved physico-chemical properties. These properties enhance the sensitivity of the sensors, which can also be used to open up new applications such as nanostructure transistors and $n$- or $p$-type nanoparticle-functionalization with bio-molecules for medical diagnosis.       

\section*{Acknowledgments}

This work was supported by the Slovene Human Resources Development and Scholarship Fund (Ad-Futura), the Slovenian Research Agency (ARRS) and the Institut Jo$\check{\rm z}$ef Stefan (IJS). We thank Matev$\check{\rm z}$ Pi$\check{\rm c}$man for his contribution to develop an animation given in the supplementary information. A.D.A. gratefully acknowledge the financial support provided by the late Madam Kithriammal Soosay who passed away suddenly while this article was in preparation. Her unconditional and continuous support during the period 1999-2010 was paramount to the development of the ionization energy theory.


\begin{thebibliography}{plain}

\bibitem{gala} K. Galatsis, L. Cukrov, W. Wlodarski, P. McCormick, K. Kalantar-zadeh, E. Comini, G. Sberveglieri, \textit{Sens. Actuators B}, \textbf{2003}, \textit{93}, 562-565.

\bibitem{biel} A. Bielanski, J. Deren, J. Haber, \textit{Nature}, \textbf{1957}, \textit{179}, 668-679.

\bibitem{lee} Y. C. Lee, Y. L. Chueh, C. H. Hsieh, M. T. Chang, L. J. Chou, Z. L. Wang, Y. W. Lan, C, D. Chen, H. Kurata, S. Isoda, \textit{Small}, \textbf{2007}, \textit{3}, 1356-1361.

\bibitem{gur} A. Gurlo, N. Barsan, A. Oprea, M. Sahm, T. Sahm, U. Weimar, \textit{Appl. Phys. Lett.}, \textbf{2004}, \textit{85}, 2280-2282.

\bibitem{ash} N. W. Ashcroft, N. D. Mermin, \textit{Solid State Physics}, Thomson Learning, Inc., Australia, \textbf{1976}.

\bibitem{gouma} P. I. Gouma, A. K. Prasad, K. K. Iyer, \textit{Nanotechnology}, \textbf{2006}, \textit{17}, S48-S53.

\bibitem{pras} A. K. Prasad, P. I. Gouma, \textit{J. Mater. Sci.}, \textbf{2003}, \textit{38}, 4347-4352.

\bibitem{uros} U. Cvelbar, K. Ostrikov, A. Drenik, M. Mozeti$\check{\rm c}$, \textit{Appl. Phys. Lett.}, \textbf{2008}, \textit{92}, 133505-133507.

\bibitem{gouma2} P. I. Gouma, K. Kalyanasundaram, X. Yun, M. Stanacevic, L. Wang, \textit{IEEE Sens. J.}, \textbf{2010}, \textit{10}, 49-53.

\bibitem{sunu} S. S. Sunu, E. Prabhu, V. Jayaraman, K. I. Gnanasekar, T. Gnanasekaran, \textit{Sens. Actuators B}, \textbf{2003}, \textit{94}, 189-196.

\bibitem{chau} S. Chauhana, S. Chauhanb, R. D'Cruzf, S. Faruqic, K. K. Singhd, S. Varmae, M. Singha, V. Karthik, \textit{Environ. Toxicol. Pharmacol.}, \textbf{2008}, \textit{26}, 113-122.

\bibitem{bars} N. Barsan, C. Simion, T. Heine, S. Pokhrel, \textit{J. Electroceram.}, \textbf{2010}, \textit{133}, 11-19.

\bibitem{smit} E. de Smit, I. Swart, J. F. Creemer, G. H. Hoveling, M. K. Gilles, T. Tyliszczak, P. J. Kooyman, H. W. Zandbergen, C. Morin, B. M. Weckhuysen, F. M. F. de Groot, \textit{Nature}, \textbf{2008}, \textit{456}, 222-225.

\bibitem{kaur} J. Kaur, V. D. Vankar, M. C. Bhatnagar, \textit{Sens. Actuators B}, \textbf{2008}, \textit{133}, 650-655.

\bibitem{chir} M. Chirita, I. Grozescu, \textit{Chem. Bull.}, (Romania) Politehnica Univ. (Timisoara), \textbf{2009}, \textit{54}, 1-8.

\bibitem{soo} S. C. Lee, H. Y. Choi, S. J. Lee, W. S. Lee, J. S. Huh, D. D. Lee, J. C. Kim, \textit{Sens. Actuators B}, \textbf{2009}, \textit{138}, 446-452.

\bibitem{cwang} C. Wang, J. Chen, T. Talavage, J. Irudayaraj, \textit{Angew. Chem. Int. Ed.}, \textbf{2009}, \textit{48}, 2759-2763.

\bibitem{mohr} R. Mohr, K. Kratz, T. Weigel, M. Lucka-Gabor, M. Moneke, A. Lendlein, \textit{Proc. Natl. Acad. Sci.}, \textbf{2006}, \textit{103}, 3540-3545.

\bibitem{nair} G. Nair, S. M. Geyer, L. -Y. Chang, M. G. Bawendi, \textit{Phys. Rev. B}, \textbf{2008}, \textit{78}, 125325-125334.

\bibitem{nair2} G. Nair, M. G. Bawendi, \textit{Phys. Rev. B}, \textbf{2007}, \textit{76}, 081304(R)-081307(R).

\bibitem{hua} H. Cao, G. Wang, L. Zhang, Y. Liang, S. Zhang, X. Zhang, \textit{ChemPhysChem}, \textbf{2006}, \textit{7}, 1897-1901.

\bibitem{xie} J. Xie, J. Wang, G. Niu, J. Huang, K. Chen, X. Li, X. Chen, \textit{Chem. Commun.}, \textbf{2010}, \textit{46}, 433-435.

\bibitem{huber} D. L. Huber, \textit{Small}, \textbf{2005}, \textit{1}, 482-501.

\bibitem{and1} a) A. D. Arulsamy, \textit{Physica C}, \textbf{2001}, \textit{356}, 62-66; b) A. D. Arulsamy, \textit{Phys. Lett. A}, \textbf{2002}, \textit{300}, 691-696.

\bibitem{and2} A. D. Arulsamy, X. Y. Cui, C. Stampfl, K. Ratnavelu, \textit{Phys. Status Solidi B}, \textbf{2009}, \textit{246}, 1060-1071. 

\bibitem{po} P. C. Chen, S. C. Mwakwari, A. K. Oyelere, \textit{Nanotechnology, Science and Applications}, \textbf{2008}, \textit{1}, 45-66. 

\bibitem{gup} a) A. K. Gupta, M. Gupta, \textit{Biomaterials}, \textbf{2005}, \textit{26}, 3995-4021; b) A. K. Gupta, M. Gupta, \textit{Biomaterials}, \textbf{2005}, \textit{26}, 1565-1573. 

\bibitem{guo} M. Guo, Y. Yan, X. Liu, H. Yan, K. Liu, H. Zhang, Y. Cao, \textit{Nanoscale}, \textbf{2010}, \textit{2}, 434-441. 

\bibitem{rui} R. Qiao, C. Yang, M. Gao, \textit{J. Mater. Chem.}, \textbf{2009}, \textit{19}, 6274-6293. 

\bibitem{chin} C. Fang, M. Zhang, \textit{J. Mater. Chem.}, \textbf{2009}, \textit{19}, 6258-6266. 

\bibitem{jin} J. Park, M. K. Yu, Y. Y. Jeong, J. W. Kim, K. Lee, V. N. Phan, S. Jon, \textit{J. Mater. Chem.}, \textbf{2009}, \textit{19}, 6412-6417. 

\bibitem{ver} J. E. W. Verwey, P. W. Haayman, \textit{Physica}, \textbf{1941}, \textit{8}, 979-987. 

\bibitem{chen} Z. Chen, U. Cvelbar, M. Mozeti$\check{\rm c}$, J. He, M. K. Sunkara, \textit{Chem. Mater.}, \textbf{2008}, \textit{20}, 3224-3228.

\bibitem{uros2} U. Cvelbar, Z. Chen, M. K. Sunkara, M. Mozeti$\check{\rm c}$, \textit{Small}, \textbf{2008}, \textit{4}, 1610-1614.

\bibitem{moz} M. Mozeti$\check{\rm c}$, U. Cvelbar, M. K. Sunkara, S. Vaddiraju, \textit{Adv. Mater.}, \textbf{2005}, \textit{17}, 2138-2142.

\bibitem{vai} B. Vaidhyanathan, S. Murugavel, S. Asokan, K. J. Rao, \textit{J. Phys. Chem. B}, \textbf{1997}, \textit{101}, 9717-9726.

\bibitem{muru} S. Murugavel, S. Asokan, \textit{Phys. Rev. B}, \textbf{1998}, \textit{58}, 4449-4453.

\bibitem{kolo} A. V. Kolobov, \textit{J. Non-Cryst. Solids}, \textbf{1996}, \textit{198-200}, 728-731.

\bibitem{ell} S. R. Elliot, A. T. Steel, \textit{Phys. Rev. Lett.}, \textbf{1986}, \textit{57}, 1316-1319.

\bibitem{and3} A. D. Arulsamy, \textit{Pramana J. Phys.}, \textbf{2010}, \textit{74}, 615-631.

\bibitem{and4} A. D. Arulsamy, PhD thesis, The University of Sydney (Australia), \textbf{2009}.

\bibitem{and8} a) A. D. Arulsamy, \textit{Phys. Lett. A}, \textbf{2005}, \textit{334}, 413-421; b) A. D. Arulsamy, K. Ostrikov, \textit{Phys. Lett. A}, \textbf{2009}, \textit{373}, 2267-2272.

\bibitem{anddd} a) A. D. Arulsamy, K. Ostrikov, \textit{Physica B}, \textbf{2010}, \textit{405}, 2263-2271; b) A. D. Arulsamy, U. Cvelbar, M. Mozeti$\check{\rm c}$, K. Ostrikov, \textit{Nanoscale}, \textbf{2010}, \textit{2}, 728-733.

\bibitem{web} M. J. Winter, $\langle$www.webelements.com$\rangle$. The Elements Periodic Table: Essential Data and Description.

\bibitem{band} M. B. Sahana, C. Sudakar, G. Setzler, A. Dixit, J. S. Thakur, G. Lawes, R. Naik, V. M. Naik, P. P. Vaishnava, \textit{Appl. Phys. Lett.}, \textbf{2008}, \textit{93}, 231909-231911.

\bibitem{band2} H. K. Bowen, D. Adler, B. H. Auker, \textit{J. Solid State Chem.}, \textbf{1975}, \textit{12}, 355-359.

\bibitem{marco} A. D. Arulsamy, M. Fronzi, \textit{Physica E}, \textbf{2008}, \textit{41}, 74-79.

\bibitem{oxide} A. D. Arulsamy, K. Eler$\check{\rm s}$i$\check{\rm c}$, M. Modic, U. Cvelbar, M. Mozeti$\check{\rm c}$, Renormalized ionic polarizability functional applied to evaluate the molecule-oxide interactions (In preparation: arXiv:1003.4625).  

\bibitem{reed} T. M. Reed, \textit{J. Phys. Chem.}, \textbf{1955}, \textit{59}, 428-432.

\bibitem{bog} M. P. Bogaard, A. D. Buckingham, R. K. Pierens, A. H. White, \textit{J. Chem. Soc. Faraday Trans.}, \textbf{1978}, \textit{74}, 3008-3015.

\bibitem{andrew} A. D. Arulsamy, Renormalization group method based on the ionization energy theory (In preparation: arXiv:0807.0745).  

\bibitem{gurlo} A. Gurlo, R. Riedel, \textit{Angew. Chem. Int. Ed.}, \textbf{2007}, \textit{46}, 3826-3848.

\bibitem{ch} Y. L. Chueh, M. W. Lai, J. Q. Liang, L. J. Chou, Z. L. Wang, \textit{Adv. Funct. Mater.}, \textbf{2006}, \textit{16}, 2243-2251.

\bibitem{jas} J. Jasinski, K. E. Pinkerton, I. M. Kennedy, V. J. Leppert, \textit{Sens. Actuators B}, \textbf{2005}, \textit{109}, 19-23.

\bibitem{fan} Z. Fan, X. Wen, S. Yang, J. G. Lu, \textit{Appl. Phys. Lett.}, \textbf{2005}, \textit{7}, 013113-013115.

\bibitem{gurr2} A. Gurlo, M. Sahm, A. Oprea, N. Barsan, U. Weimar, \textit{Sens. Actuators B}, \textbf{2004}, \textit{85}, 291-298.

\bibitem{gurlo2} A. Gurlo, \textit{ChemPhysChem}, \textbf{2006}, \textit{7}, 2041-2052.

\end{thebibliography}
\end{document}